\documentclass{elsart}

\usepackage{graphicx}

\usepackage{amssymb}

\newcommand{\ap}{\mbox{$\overline{{\rm p}}$ }}
\newcommand{\el}{\mbox{${\rm e^{-}}$ }}
\newcommand{\ps}{\mbox{${\rm e^{+}}$ }}

\begin{document}
 
\begin{frontmatter}

\title{A High Granularity Imaging Calorimeter for Cosmic-Ray Physics}

\author{M. Boezio\corauthref{cor}},
\corauth[cor]{Corresponding author. Tel.: +39-040-3756224; 
fax: +39-040-3756258}
\ead{Mirko.Boezio@trieste.infn.it}
\author{V. Bonvicini},
\author{E. Mocchiutti\thanksref{emiliano}},
\thanks[emiliano]{Now at Royal Institute of Technology, Stockholm, Sweden}
\author{P. Schiavon},
\author{G. Scian},
\author{A. Vacchi},
\author{G. Zampa},
\author{N. Zampa}
\address{INFN Section and Physics Department, University of Trieste, 
Via A. Valerio 2, 34127 Trieste, Italy}

\begin{abstract}
An imaging calorimeter has been designed 
and is being built for the PAMELA satellite-borne experiment.
The physics goals of the experiment are the measurement of the 
flux of antiprotons, positrons and light isotopes in the cosmic radiation.

The calorimeter is designed 
to perform a precise measurement of the total energy deposited, to 
reconstruct the spatial development of the showers (both 
in the longitudinal and in the transverse directions), and to measure the 
energy distribution along the shower itself. From this information,
the calorimeter will identify antiprotons
from a electron background 
and positrons in a background of protons with an efficiency of 
about 95\% and a rejection power better than 10$^{-4}$. 
Furthermore, a self-trigger system has been implemented with the 
calorimeter that will be employed to measure high-energy 
(from about 300~GeV to more than 1~TeV) electrons.

The instrument is composed of 22 layers of tungsten, 
each sandwiched between two ``views" of 
silicon strip detectors (X and Y).  
The signals are read out by a custom VLSI front-end 
chip, the CR1.4P, specifically designed for the PAMELA 
calorimeter, with a dynamic range of 
7.14 pC or 1400 mip (minimum ionizing particle).

We report on the simulated 
performance and prototype design.

\end{abstract}


\begin{center}
{\bf Submitted to Nuclear Instruments \& Methods in Physics Research}
\end{center}

\end{frontmatter}

\clearpage

\section{Introduction: motivation for an Imaging Calorimeter in PAMELA}

The PAMELA experiment is part of the Russian Italian Mission (RIM) 
program, which foresees several space missions with different scientific 
objectives. Two missions have already been carried out. 
The first one \cite{bid00} studied the anomalous
light flashes perceived by astronauts in orbit by employing silicon detectors
placed around the astronaut's head and it was conducted on-board the
space station MIR (RIM-0 experiment). The other studied the low energy 
isotopic composition of cosmic nuclei by means of the silicon telescope
NINA \cite{bak97}, carried by the Russian polar orbit 
satellite Resource-04 (RIM-1 experiment). The 
NINA instrument was launched successfully from Bajkonur on July 10th, 1998 
and has collected a significant amount of data \cite{bid01}.

The PAMELA \cite{bon00} experiment (RIM-2 mission) has the scientific goal of 
measuring the cosmic radiation over a wide energy range. The PAMELA 
apparatus will be installed on board of a Russian satellite of the Resource 
series and will be launched in early 2003. Its sun-synchronous, 600~km 
nearly polar orbit will allow  the low energy cosmic rays to be measured while 
the instrument is near the poles. The main objectives of the 
experiment are the precise measurement of the positron flux from 50~MeV to 
270~GeV and the antiproton flux from 80~MeV to 190~GeV, as well as the 
search for anti-helium with a sensitivity of $10^{-7}$ in the 
{\mbox{$\overline{{\rm He}}$ }}/He ratio. 
For further details see \cite{bon00}.

\begin{figure}[ht]
\begin{center}
\includegraphics[height=100mm]{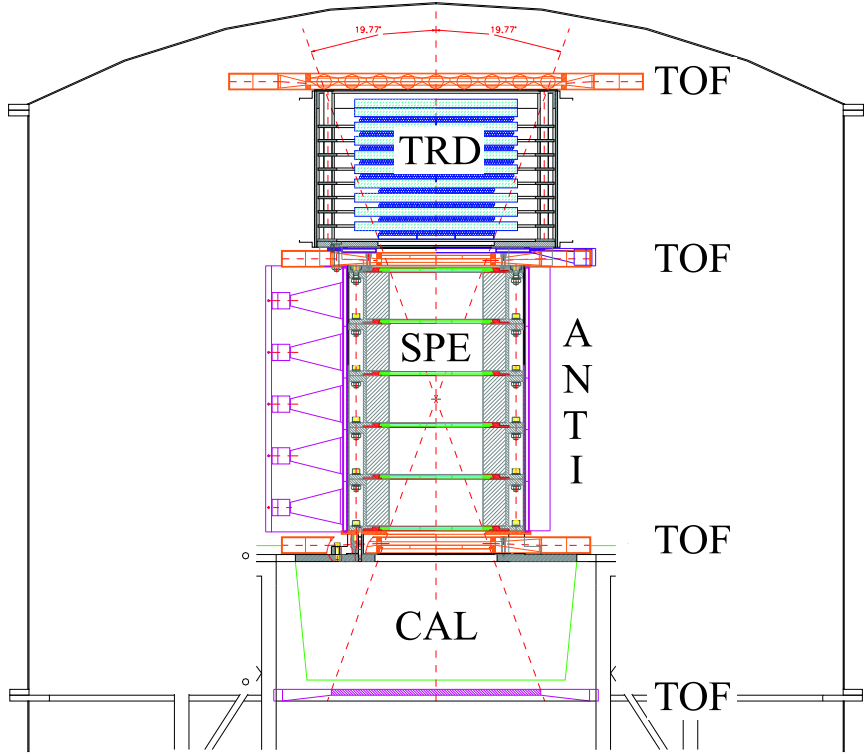}
\end{center}
\caption[]{The PAMELA telescope
\label{pamela}}
\end{figure}

PAMELA has a detector layout similar to the one used by the WiZard 
Collaboration in its balloon experiments \cite{amb99}. 
The apparatus is composed of 
the following subdetectors, arranged as in fig.~\ref{pamela}:
\begin{itemize}
  \item a plastic scintillator system that includes a time of 
flight counter 
(TOF), which provides energy loss, timing information and the
trigger for the data acquisition system, 
 and an anti-coincidence system (ANTI), which identifies those particles 
entering the spectrometer from outside its geometrical acceptance;
 \item a Transition Radiation Detector (TRD), which provides a threshold 
velocity 
measurement, complementing the calorimeter in the particle identification;
\item a magnetic spectrometer (SPE), formed by a permanent magnet 
(supplying a field of 0.4~T) and a tracking system 
equipped with 6 layers of double-sided silicon microstrip 
detectors for precise track reconstruction, capable of determining the sign 
and the absolute value of the electric charge with a very high confidence 
level and measuring the momentum of the particles up to the highest energies;
 \item an electromagnetic imaging calorimeter (CAL), which  measures the 
energy 
released by the interacting electrons and positrons and reconstructs the 
spatial development of the shower, allowing to distinguish electromagnetic 
showers from hadronic showers and from non-interacting particles;
\item a plastic scintillator counter mounted under the calorimeter
for triggering of high energy ($>$~100~GeV) electrons.
\end{itemize}
The total height of the apparatus is 123.6~cm and the lateral dimensions of 
the detectors have been determined in such a way to fully cover the 
acceptance of the magnet spectrometer.
Furthermore, a neutron counter is foreseen to be installed in the payload
along with the PAMELA apparatus and just below the calorimeter. This 
additional detector will work together with the calorimeter to measure
very-high-energy electrons (see section~\ref{self}).

The apparatus has been carefully designed taking into account the strict 
mass (430~kg) and power consumption (350~W) limitations imposed by the 
satellite environment as well as the need for mechanical stability to 
withstand the vibration and shock loads occurring during the launch phase.
In the following, we present the prototype design 
and the simulated performance.

\section{General characteristics of the calorimeter}

The PAMELA Imaging Calorimeter is a sampling calorimeter made of silicon 
sensor planes interleaved with plates of tungsten absorber. The application 
of silicon detectors as active layers for sampling calorimeters is a well 
established technique in experimental high-energy physics \cite{pen88}. 
Among the advantages of these detectors are excellent stability, linearity, 
efficiency, low-voltage operation. These features allow to extend their use 
also to cosmic-ray experiments on satellites.

The instrument was designed aiming to a high segmentation, both in the 
longitudinal (Z) and in the transversal (X and Y) directions. In the Z 
direction, the granularity is determined by the thickness of the layers of 
absorbing material. Each tungsten layer has a thickness of 0.26~cm, which 
corresponds to 0.74~X$_{0}$ (radiation lengths). Since there are 22 tungsten 
layers, the total depth is 16.3~X$_{0}$ (i.e. about 0.6 interaction lengths). 
The depth of the instrument is not sufficient to fully contain high-energy 
electromagnetic showers. However, the granularity, along with the 
energy resolution of the silicon detectors, allows an accurate topological
reconstruction of the shower development, making the calorimeter a 
powerful particle identifier, as confirmed by simulation results (see 
section~\ref{s:sim}).
The transverse granularity is given by the segmentation of the silicon 
detectors into strips. The silicon detectors for the PAMELA calorimeter are 
large area devices ($8 \times 8$ cm$^{2}$ each), 
380~$\mu$m thick and segmented into 32 large 
strips with a pitch of 2.4~mm. They are fabricated on high-purity, 
high-resistivity ($\ge 7 k \Omega \cdot$cm), 
n-type silicon substrate. In order to avoid 
the use of a conductive epoxy glue for gluing the detectors onto the printed 
circuit boards (and therefore protecting the devices from the  
mechanical stress that this type of glues can induce during polymerization), 
a special design allows bringing the bias voltage directly on the junction 
side of the devices via wire bonding.
 
Each tungsten plane is sandwiched between two layers of silicon detectors, 
i.e. the layout of a single plane is Si-X/W/Si-Y. Either type of view 
(X or Y) is made by nine silicon detectors, arranged in a square matrix of 
$3 \times 3$ detectors. The total sensitive area is about 
$24 \times 24$~cm$^{2}$. Each of the 32 
strips of a detector is connected to those belonging to the other 
two detectors 
of the same row (or column) forming 24~cm long strips.
The number of electronics channels per 
plane is $32 \times 3 \times 2 = 192$ and the total number of channels 
is $192 \times 22 = 4224$.

\section{Mechanical structure and assembly techniques}

\begin{figure}[ht]
\begin{center}
\includegraphics[height=100mm]{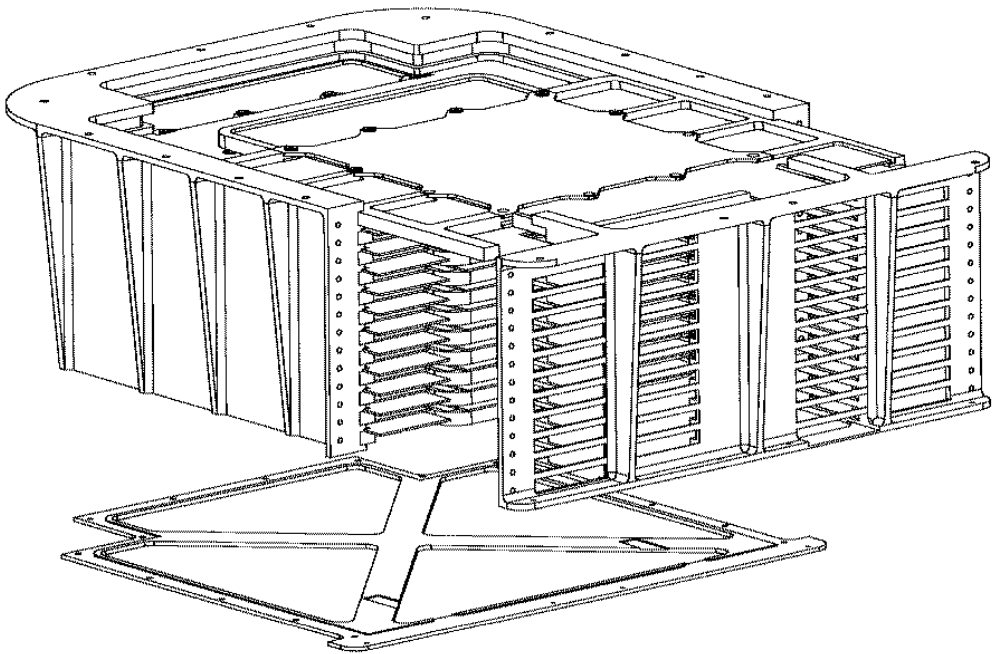}
\end{center}
\caption[]{Schematic of the calorimeter mechanical structure
\label{calo}}
\end{figure}

In order to be qualified for a space flight, the mechanical structure of the 
calorimeter (as for any other PAMELA sub-detector) must be able to withstand 
the vibrations and shocks occurring during the launch phase. Given the above 
mentioned constraints on the available mass budget, the calorimeter mechanics 
has been designed aiming at a robust and light structure, allowing simple 
operations and easy access to the single planes. 
On the basis of careful simulations, we chose a 
special, space-qualified aluminium alloy, since it allowed to have 
a light, robust and cheap solution.
The mechanical structure was designed on the basis of a modular concept. 
The basic unit is called a "detection plane", and it consists of an absorber 
plate (made by a special sintered tungsten alloy, 2.6~mm thick), two 
multi-layer printed circuit boards (for X and Y views, supporting the 
silicon detectors and the front-end electronics up to the ADC and called 
"FE boards") and the two matrices of silicon sensors. Two such detection 
planes form a "detection module". In a module, the two detection planes 
are kept together by special aluminium frames to which they are bolted 
at the edge of the absorber plates. All modules are independent and 
fully extractable; they are inserted like ``drawers" into the mainframe 
through precisely machined guides and then locked in place by a cover 
(Fig.~\ref{calo}).
As previously described, there are 22 tungsten plates in the calorimeter, 
which correspond to 11 modules. There is also a 12th module, formed 
by two aluminium ``dummy plates", that has the sole purpose of supporting 
the four read-out boards (see next section). The total calorimeter mass 
is about 110~kg.
 
\begin{figure}[ht]
\begin{center}
\includegraphics[height=80mm]{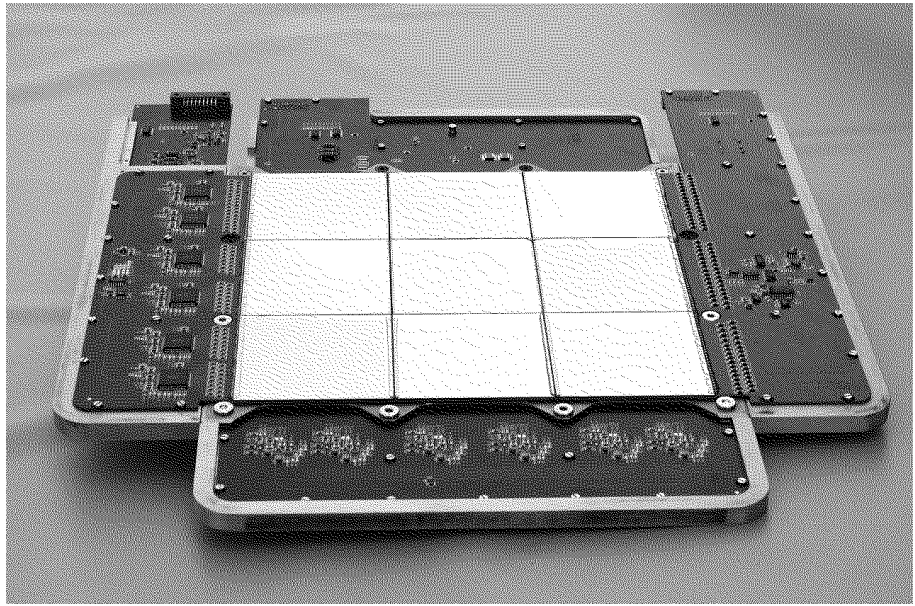}
\end{center}
\caption[]{Front-end board.
\label{frontend}}
\end{figure}

The interconnection and assembly technique for the detection planes has been 
defined and optimised through numerous tests\footnote{In collaboration 
with the firm 
MIPOT (Cormons, Italy, http://www.mipot.com/).}. 
The silicon detectors are glued in rows of three on a 
specially designed, 50~$\mu$m thick kapton layer using a siliconic glue. Then, 
the wire bonding of the corresponding strips on each detector is performed. 
This realises a ``ladder". The wire bondings are then coated by a deposition 
of a layer of siliconic glue. As a next step, the FE boards are first 
equipped with all electronic components by using an automatic placing and 
soldering procedure, and then fixed to the corresponding tungsten plates by 
means of a silicon glue. Afterwards, three ladders are glued onto each FE 
circuit board, again by using a special siliconic glue, to form the 
$3 \times 3$ 
silicon detector matrix of either view. Finally, the wire 
bonding of the strips to the preamplifiers is performed. Fig.~\ref{frontend} 
shows a 
complete front-end board mounted in a module.

\section{Signal processing and data acquisition electronics}

\subsection{Front-end electronics}
\begin{figure}[ht]
\begin{center}
\includegraphics[height=100mm]{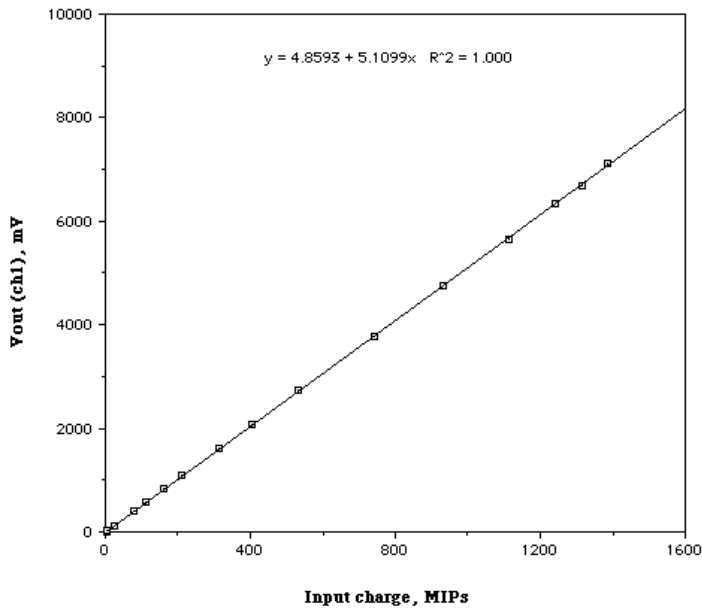}
\end{center}
\caption[]{Measured linear range in one channel of the CR1.4P chip. 
\label{lincheap}}
\end{figure}

The front-end electronics is based on a VLSI ASIC specifically designed 
for the PAMELA calorimeter: the CR1.4P \cite{ada99}. 
The final version of the chip 
was attained after an R\&D phase that went through the production and 
characterisation of three different prototypes. The use of an ASIC for 
processing the analog signals from the silicon detectors allows considerable 
weight saving and design compactness with respect to the discrete 
preamplifiers previously used in the calorimeters for balloon 
flights \cite{boc96}.
The main design characteristics of this chip, which is fabricated in a 
CMOS 2~$\mu$m mixed analog/digital technology, are the wide dynamic range 
(1400 minimum ionising particles, $1~{\rm mip} \simeq 5.1$~fC for 
380~$\mu$m thick silicon 
detectors), the ability to cope with a large ($\simeq 180$ pF) detector 
capacitance, the good noise performance (ENC $\simeq$ 2700 e$^{-}$ rms + 
5 e$^{-}$/pF) and 
the low power consumption (6~mW/channel). 
Fig.~\ref{lincheap} shows the measured linear 
range in one channel of the chip. Over the full range, the maximum deviation 
from the fitted line is $<$ 2.5\% and the average linearity is better than 1\%.
One circuit has 16 channels, each comprising a charge sensitive 
preamplifier (CSA), a shaping amplifier/filter, a track-and-hold circuit 
and an output multiplexer. A self-trigger system and an input calibration 
circuit are also integrated on chip. 
\begin{figure}[ht]
\begin{center}
\includegraphics[height=70mm]{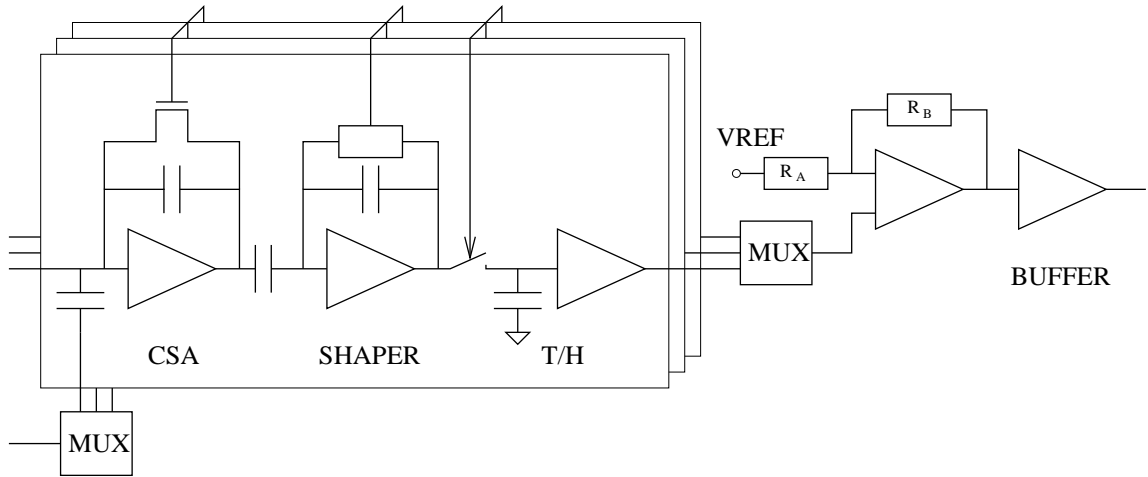}
\end{center}
\caption[]{Architecture of a single channel of the CR1.4P chip.
\label{archch}}
\end{figure}
Fig.~\ref{archch} shows the general architecture 
of a single channel. The CSA, based on a folded-cascode architecture, 

integrates the charge from the silicon detector. An MOS transistor is used 
in the feedback loop to reset the charge from the integrating capacitor. 
The output of the CSA is fed into a CR-RC shaping amplifier. The shaper 
peaking time can be adjusted from about 800~ns to 1.2~$\mu$s by acting on the 
shaper bias voltage and current. In such a way, the signal-to-noise ratio 
can be optimised (within a certain range) for a specific application. In 
PAMELA, we use a peaking time of 1.2~$\mu$s. A track-and-hold (T/H) circuit 
follows the shaper and stores the peak value of the signal. The control of 
the T/H switch is done with a differential signal. A 
multiplexer (MUX), formed by 
a 16-bit shift register that controls an array of analog switches, 
connects one analog channel at a time to the output stage. This is composed 
by an amplifier with dc-biased output and by an output buffer. Another 
shift register allows to apply a calibration signal to any of the inputs 
(or any pattern of them) through a 2~pF injection capacitor. Therefore, 
by performing periodic calibration runs, 
it is possible to monitor the behaviour of the device and its stability 
after the launch and during its operation in space. 

\begin{figure}[ht]
\begin{center}
\includegraphics[height=80mm]{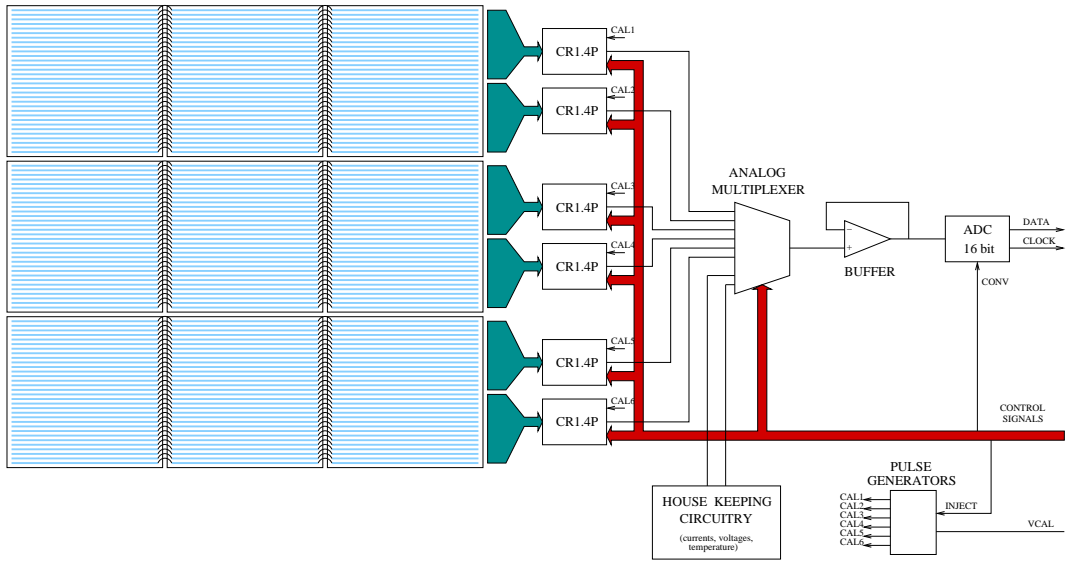}
\end{center}
\caption[]{Block scheme of the FE board.
\label{bsFE}}
\end{figure}

The high energy resolution needed to identify minimum ionizing particles 
demands a 16 bit analog to digital converter.
Such ADCs are usually slow and require careful board design to
be able to exploit their full potential. For these reasons it
has been decided to place them as close as possible to the CR1.4P
chips on the FE board (a block scheme of which is depicted in
Fig.~\ref{bsFE}). To separate the data paths, 
thus minimizing the possible failure points, 
an ADC converter with a serial interface (AD977A by Analog Devices
Inc.) has been chosen.
Among the available ADC's voltage ranges, the most suited one is
the 6.66V, which maximizes the energy resolution at an acceptable
dynamic range loss of about 18\% (1150~mip instead of 1400~mip).
The six CR1.4P are connected to the ADC through an 8:1 multiplexer
and a unity-gain buffer which drives the relatively low input
impedance of the converter (about 5~K$\Omega$).
The two other inputs of the multiplexer can be advantageously
used to acquire some useful electrical quantities (so-called
House Keepings, HK) that allow to monitor the functionality of
the calorimeter.
The first of this channels measures the current of the detector's
active area, while the other is connected to one of the other
signals (temperature, calibration voltage, analog supply voltages).
Each FE board has a sensor which provides an alarm when the
temperature rises above a given threshold: its output is
open collector, so that several alarms can be combined to simplify
the circuitry. Then the alarms are fed to the CPU where appropriate
actions can be taken.
The interface to the read-out has been realized through digital
signals arranged into an input control bus, connected in parallel
to all the FE boards, and the ADC's output signals (data
and clock). A 10 K$\Omega$ resistor has been placed in series at
each input to limit the bus load in the event of a component's
failure that would otherwise block the FE controls. A 22 pF
capacitor is connected in parallel to these resistors to pass the
signal edges. The same circuitry has been implemented on the
CR1.4P control bus.

\subsection{Read-out electronics}
The tasks of the read-out electronics are to collect and analyse
the data prior to their transmission to the main CPU.
Four read-out boards are housed in the 12th, and last, ``dummy'' calorimeter
module.

The read-out boards are connected to the front-end electronics by
means of four bus boards: this solution exploits the mechanical
design to divide the calorimeter into four independent sections
(odd X-views, odd Y-views, even X-views and even Y-views),
each one read out by one board.
Each read-out board consists of a micro-controller dedicated to the
``slow control" of the calorimeter, an ADSP2181 (Analog Devices Inc.) 
which controls the
acquisition sequence and elaborates the data, and an FPGA to parallelise
the ADC data and generate some of the control signals.

Inside the FPGA a pipeline memory has been realized. It consists
of several shift registers that parallelise the ADC data (``horizontal
shift") connected together to form a second level shifter which
brings the data to the DSP (``vertical shift").
The shifting direction is selected by the DSP: horizontal during
conversion, vertical otherwise. To account for a possible loss
of performance of the ADC due to radiation effects, the Least
Significant Bit (LSB) of the conversion is discharged.

The micro-controller is connected to the main CPU via a low
speed serial link and its tasks are to bootstrap the DSP, control
its activity by means of a ``watchdog'' mechanism, and monitor the
temperature alarms. It is connected to the DSP through the IDMA
port to have access to its internal memory, providing also a backup
solution to read the event data in case of a failure in the fast
DSP link. As soon as a trigger signal is received, the DSP starts
the acquisition procedure enabling the first channel of the CR1.4P
chips, selecting the appropriate multiplexer channel and starting
a conversion. The sequence does not rely on the ADCs busy signals,
instead the DSP program has been designed to perform some
calculations during the conversion time.

At the end of a conversion, the processor reads the data
coming from the eleven FE boards, selects the next channel
to acquire, and starts a new conversion. Waiting for the
ADC to finish its task, the DSP proceeds comparing the new
data with the thresholds of the various channels in order
to perform a zero suppression, thus reducing the amount of
memory occupied by an event. Every value that exceeds
the threshold is recorded together with the address of the
channel. In the same process, the DSP calculates also the
classification variables used by the main CPU of PAMELA to
sort the events according to a pre-defined priority:
these variables are the number of strips above threshold,
the number of clusters, i.e. contiguous strips with a signal,
and the amount of energy detected in the FE board.
The threshold is determined by the pedestal value of the
channel (the output of the preamplifier with a zero input
signal) and its noise measured during the calibration of
the calorimeter.
Every data to be transmitted to the central memory
is written on a buffer: an interrupt-driven process 
periodically checks this buffer and sends any new data to the
DSP's serial link connected with the mass memory system.
Merging the frame synchronizer, which provides the start bit,
to the data line converts this synchronous link to an
asynchronous one, allowing us to avoid the transmission of
a clock signal.
The DSP also formats the event data packet adding a header
and a trailer: the first is a word that specifies the type
of acquisition (compressed event, full event or calibration)
while the last includes the number of words in the packet
and a checksum.
A full event acquisition is provided for test purposes,
in this mode the program does not perform the zero 
suppression and sends all the data without the channel addresses.
The calibration of the calorimeter is done acquiring
pedestals (1024 events) and injecting two different charges
in the preamplifier's channels (128 events for each signal)
to measure the gain in two points of the dynamic range (20\%
and 80\%). The statistic is sufficient to measure with good
accuracy these important data and also allows the determination of
the system noise (the RMS of the pedestal distributions).

The connection between the calorimeter
and the main CPU/mass memory is based on TIA/EIA-485 differential lines
whose transceivers are located on the bus boards. 

\section{Simulation}
\label{s:sim}
Similar silicon-tungsten calorimeters, differing only in layout and 
number of radiation lengths, were developed and extensively studied
by our group through simulations, beam tests \cite{boc93,boe98} 
and balloon flights \cite{gol96,boe97,boe00}. The simulated data were
compared with experimental data finding excellent agreement 
\cite{boc93,boe98,boe00}.

From these simulations, a Monte Carlo program based on the CERN 
GEANT\-/FLU\-KA-3.21 code \cite{bru94}
was developed to study the capability of the PAMELA calorimeter. 
In particular, it was studied the performance of the calorimeter concerning
the primary
scientific goals of the PAMELA experiment, that is energy reconstruction
for electrons and identification of 
positrons and antiprotons in a vast background of protons and electrons,
respectively. 

\begin{figure}[ht]
\begin{center}
\includegraphics[height=100mm]{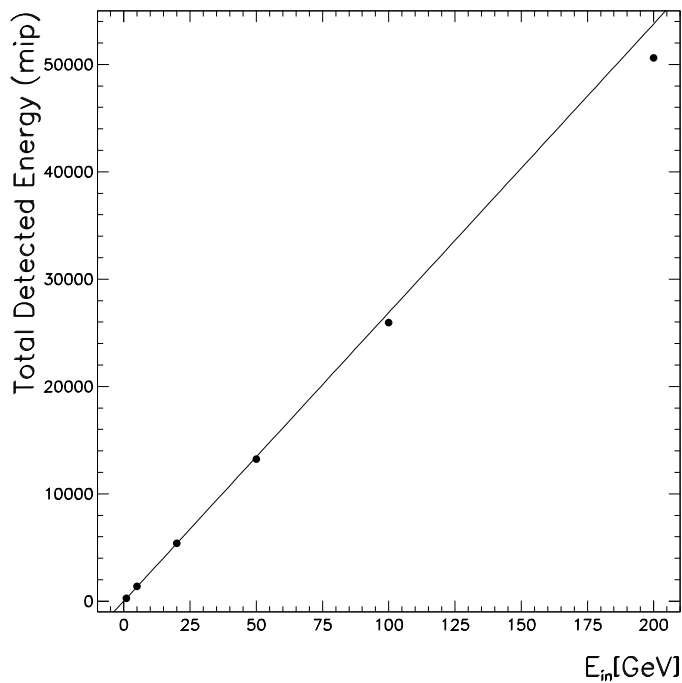}
\end{center}
\caption[]{Reconstructed energy as a function of input energy for 
simulated electrons in the PAMELA configuration. The solid line
is a linear fit to the first four points.}
\label{simlin}
\end{figure}

\begin{figure}[ht]
\begin{center}
\includegraphics[height=100mm]{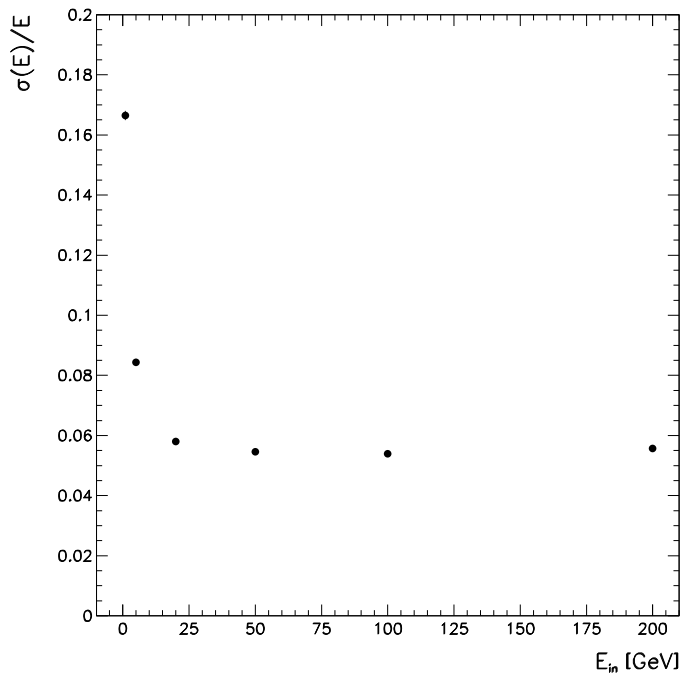}
\end{center}
\caption[]{Energy resolution as a function of input energy for simulated
electrons in the PAMELA configuration.}
\label{simres}
\end{figure}

Figure~\ref{simlin} shows the simulation of the 
total detected energy (expressed in mip units)
in the calorimeter
for electrons at several energies. The response of the calorimeter, in this 
energy range, shows a quasi-linear behaviour with deviations accounting
for the partial containment at the highest energies. The solid line is a
linear fit to the simulated data points below 100~GeV.
Figure~\ref{simres} shows the energy resolution for electrons. The 
resolution reaches a constant value above 20~GeV of about 5\%.

In a cosmic-ray experiment like PAMELA, the proton background accounts for
about 10$^{3}$ times the positron component at 1~GeV, increasing with energy
to more than 10$^{4}$ above 10~GeV. Electrons are about 10$^{2}$ more 
abundant than antiprotons at 1~GeV, decreasing with energy but still being
about 30 times more at 10~GeV. For this reason powerful particle identifiers
are needed. The PAMELA calorimeter is well suited to 
identify these particles in the cosmic radiation.

The longitudinal and transverse
   segmentation of the calorimeter combined with the measurement of the energy
   lost by the particle in each silicon strip results in a high
   identification power for electromagnetic showers. Thus, the calorimeter,
in the electron and positron analysis,
is used to identify electromagnetic showers, while in the antiproton
analysis the calorimeter is used to reject these events. 

Selection criteria were developed based on the 
following information of the electromagnetic
showers \cite{boe98}: 
\begin{enumerate}
    \item the starting point of the shower;
    \item the energy-momentum match;
    \item the longitudinal profile;
    \item the transverse profile;
    \item the topological development of the shower.
\end{enumerate}
The efficiency and contamination of the selections were 
studied simulating a large number of electrons (electrons and positrons were
assumed to be equivalent 
at the energy of interest here), antiprotons and protons. 

\begin{figure}[ht]
\begin{center}
\includegraphics[height=100mm]{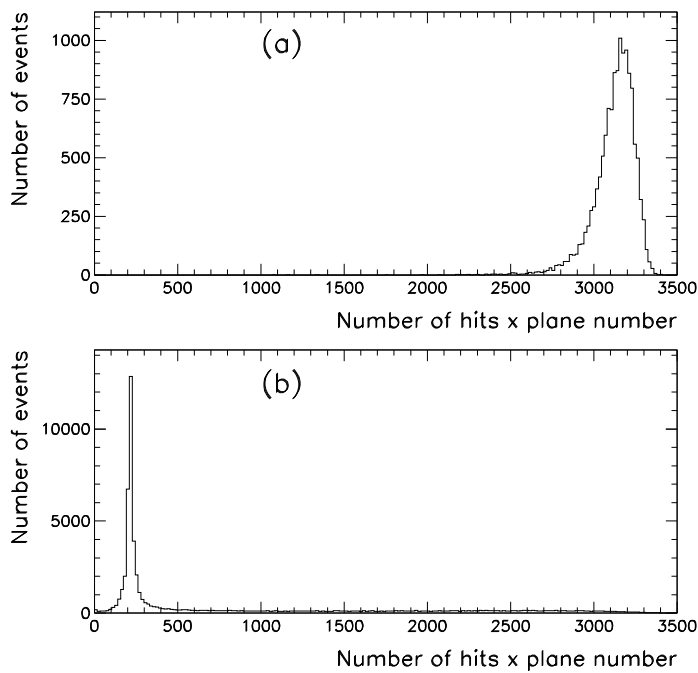}
\end{center}
\caption[]{The topological development of the shower.
The figure shows the sum of the number of hits inside 2 Moli\`{e}re units
around the track
multiplied by
the corresponding calorimeter plane number for simulated 100~GeV/c
(a) electrons and (b) protons.}
\label{sep}
\end{figure}

Figure~\ref{sep} shows 
a quantity related to the topological development of
the shower in the calorimeter for simulated
100~GeV/c (a) electrons and (b) protons . This quantity
is the product of the sum of the number of hits inside a 
cylinder, of radius about 2 Moli\`{e}re units 
(8.5 calorimeter strips) with its axis along the 
particle direction, and the corresponding 
calorimeter plane number. The direction of the
particle in the calorimeter was assumed known since in the PAMELA 
experiment it will be 
obtained from an extrapolation of the fitted
track in the tracking system.

Table~\ref{t:effcon} shows the resulting values 
for various momenta spanning the range of interest for PAMELA.

\begin{table}[t]
\caption{Simulated performances: efficiencies in antiproton and electron
detection versus electron and proton contamination, respectively.
\label{t:effcon}}
\begin{tabular}{|c|cc|cc|}
\hline
Momentum & \ap & \el & \ps & p 
\\
(GeV/c) & efficiency & contamination &  efficiency & contamination \\
\hline
1 & $0.9192 \pm  0.0009$ & $(2.5 \pm 0.2) \times 10^{-3}$ &
$0.899 \pm  0.001$ & $(1.9 \pm 0.4)\times 10^{-4}$ \\
5 & $0.9588 \pm  0.0005$ & $(4^{+5}_{-2})\times 10^{-5}$ &
$0.9533 \pm  0.0009$ & $(1.4^{+1.8}_{-0.9})\times 10^{-5}$ \\
20 & $0.9767 \pm  0.0004$ & $< 6.2 \times 10^{-5}$ &
$0.970 \pm  0.001$ & $(3^{+2}_{-1})\times 10^{-5}$ \\
100 & $0.963 \pm  0.001$ & $< 1.4 \times 10^{-4}$ &
$0.944 \pm  0.002$ & $< 3.3 \times 10^{-5}$ \\
200 & $0.954 \pm  0.002$ & $< 1.5 \times 10^{-4}$ &
$0.955 \pm  0.002$ & $< 1.2 \times 10^{-4}$ \\ 
\hline
\end{tabular}
\end{table}

\section{Self-triggering calorimeter}
\label{self}

A self-trigger system using the PAMELA 
imaging calorimeter was implemented to measure high-energy 
(from $\simeq 300$~GeV to more than 1 TeV) electrons in the 
cosmic radiation. 
Till now very few measurements have 
covered this energy range \cite{kob99}.  Since these events are quite rare in 
comparison with the normal event rate of PAMELA, it is important 
to have a large geometric factor in order to collect a reasonable statistics 
during the three year estimated lifetime of the mission. 
Requiring only that the particles enter from one of the first four planes
and cross at least
10 radiation lengths in the calorimeter, 
the overall acceptance ($\sim 600$~cm$^{2}$sr) 
becomes about a factor 30 larger than 
the normal acceptance of PAMELA.
For this reasons the CR1.4P front-end chip was designed to work not only
with an external trigger signal (``normal'' operation mode), but 
was also provided
with a ``self-trigger'' option.

\begin{figure}[ht]
\begin{center}
\includegraphics[height=75mm]{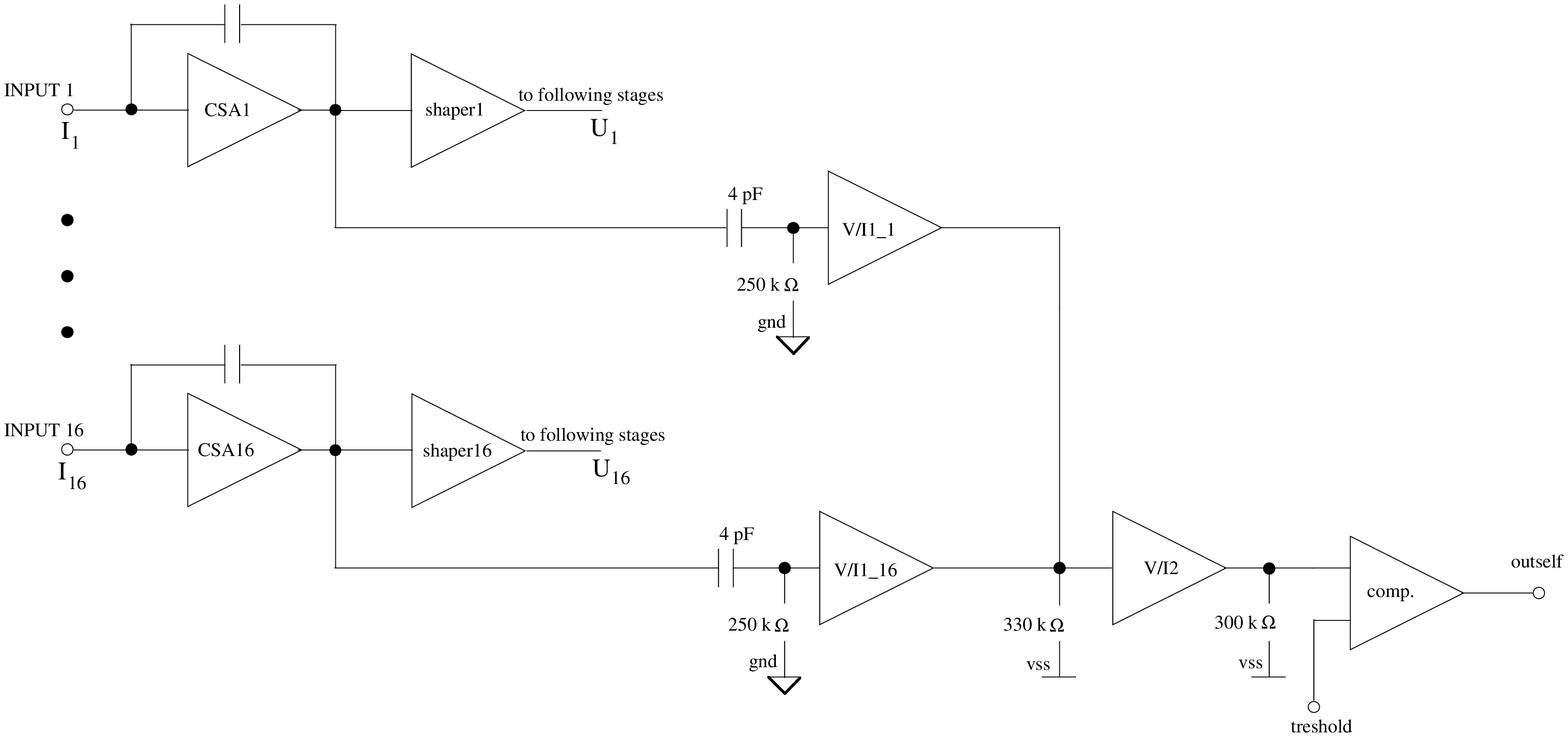}
\end{center}
\caption[]{Simplified block scheme of the self-trigger circuit in 
a CR1.4P chip.}
\label{blockscheme}
\end{figure}

\begin{figure}[ht]
\begin{center}
\includegraphics[height=100mm]{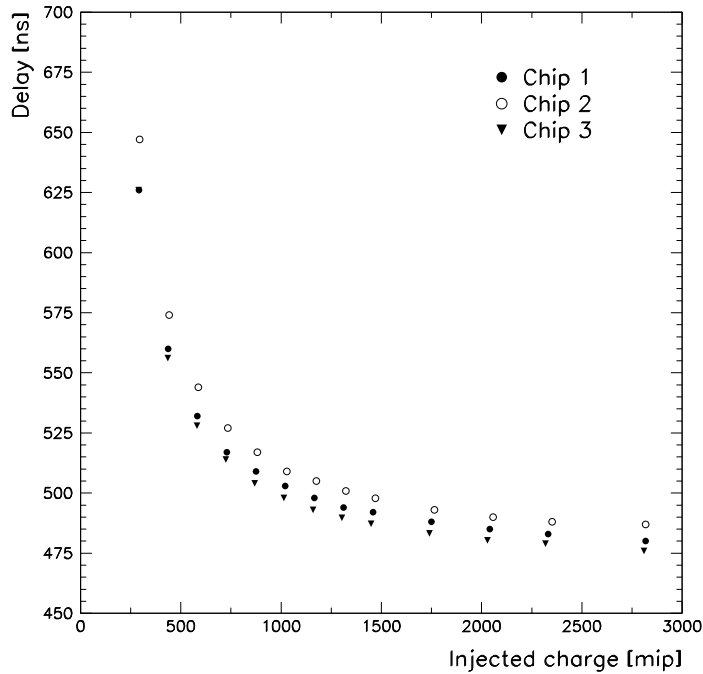}
\end{center}
\caption[]{Dependence of the self-trigger signal delay T$_{D}$ on the input 
charge for a comparator threshold of 150 mip and for three different 
chips.}
\label{delay}
\end{figure}

Figure~\ref{blockscheme} shows a simplified block scheme 
of the self-trigger circuit. 
The output of each preamplifier is transformed into a current by a 
voltage-to-current converter (V/I1\_1,...V/I1\_16). The output current of 
these stages is then summed and changed back to a voltage signal 
(across the 330 k$\Omega$ resistor). 
Another voltage-to-current converter (V/I2) 
is used to reach the desired gain ($\simeq 22$~mV/mip). A comparator gives a 
binary output signal if the sum of the signals in a chip exceeds a 
given threshold.
This kind of circuit, though performing well within its operational 
capabilities, has nonetheless some limitations, which are inherent 
in its architecture:
\begin{itemize}
	\item There is a minimum usable threshold (about 30~mip).
	\item The time delay T$_{D}$ from the injection of the input 
	charge and 
	the trigger output signal (comparator output) depends on the 
	amount of injected charge (see Figure~\ref{delay})
	and on the selected comparator threshold.
	\item Due to saturation of V/I2 output, there is an upper 
	limit to the usable threshold (about 300~mip).
\end{itemize}

The delay time is an important feature of the chip since a precise
reconstruction of the energy losses requires that the charge released in
each channel is read out at a sampling time no more than 100~ns 
different from the peaking time. 
Figure~\ref{delay} shows the delay time as a function of injected charge 
for three different chips with a 150~mip threshold. A strong dependence 
of the delay time on the input charge 
was found for low injected charge values, 
however above about 1000~mip
the delay time is essentially constant. This means that in case of high
energy losses (as it happens 
around the maximum of showers induced by high energy electrons)
a precise sampling of the energy is achievable.
It is worth noticing the small
differences, less than 20~ns, between the delay times of the three chips.

\begin{figure}[ht]
\begin{center}
\includegraphics[height=100mm]{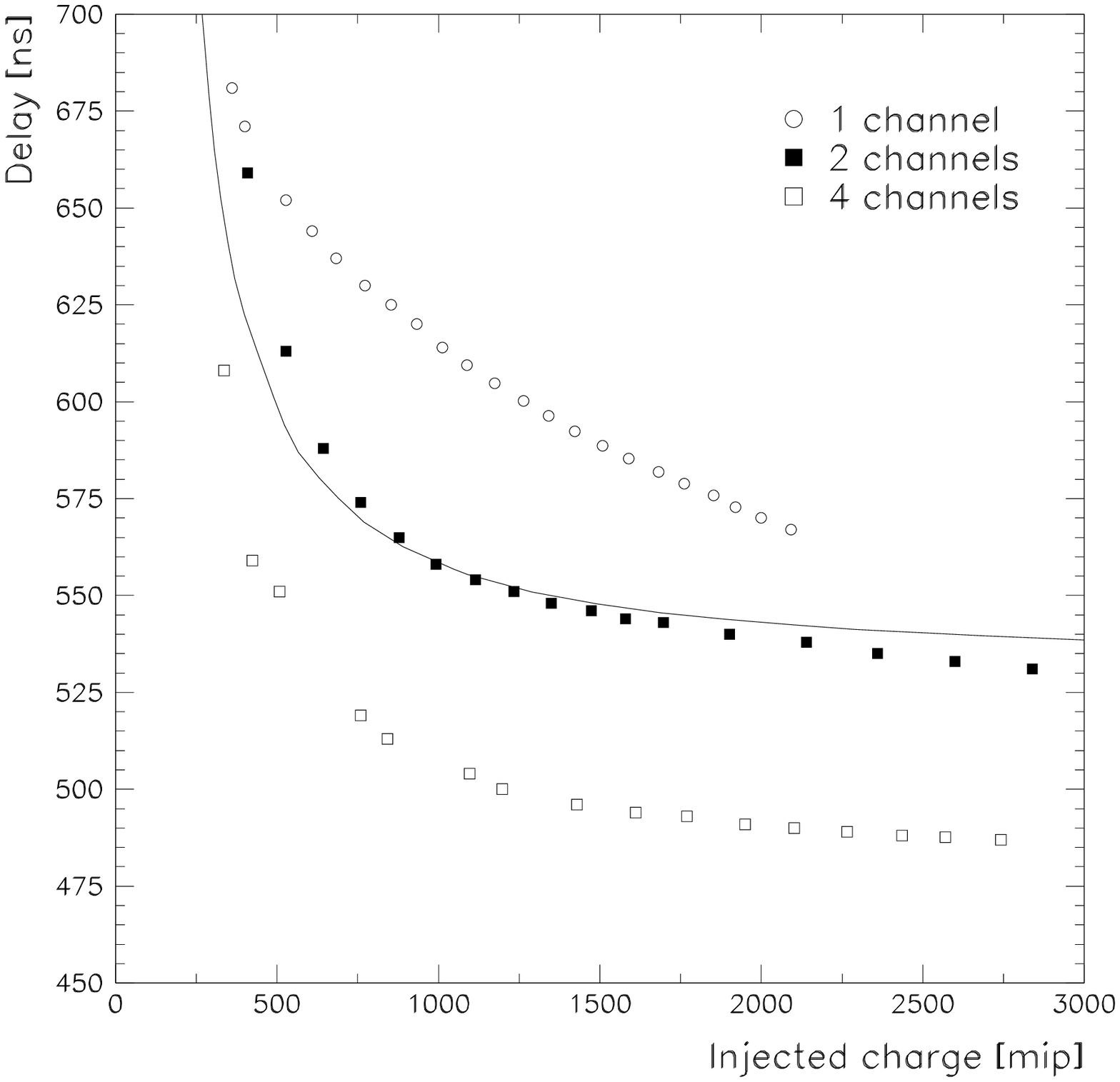}
\end{center}
\caption[]{Delay time as a function of total injected charge for a 
comparator threshold of 150 mip. The charge is injected in one (open circles),
two (full boxes) and four (open boxes) channels. The solid line 
is obtained injecting charge in the channels of the chip
according to the simulated distribution for high energy electrons.}
\label{delay2}
\end{figure}
It was found that the delay time depends also on the number of 
channels in which the charge is injected. Figure~\ref{delay2} shows the
delay time as a function of the total charge 
for a 150~mip threshold when the charge 
is injected in just one, or split in two and 
four channels (the behaviour of the 
delay time for a larger
number of channels coincides with the four channel case).
The maximum value
of the injected charge in each single channel was about 2000~mip.
It can be noticed that the asymptotic value varies with number of channels
and it differs by about 50~ns between two and four injected channels.
Furthermore, 
along with a dependence on the number of injected channels it was found 
a dependence of the delay time on how the charge was distributed among the
channels. 
Because of these effects, the variation of the delay time was 
measured injecting charge in the channels of the chip
according to the simulated distribution for electrons 
with energies greater than 300~GeV.
The best fit of these data is indicated with 
a solid line in Figure~\ref{delay2}.

As it can be seen from Figure~\ref{delay2}, the delay time 
depends on how the injected
charge is distributed among the channels 
but the differences are of the order of 50~ns. Moreover,
the behaviour of different CR chips is very similar (see Figure~\ref{delay}),
within only a few ns. 
These facts were exploited to define the self-trigger logic. 
It is worth pointing out that the trigger logic was designed 
to reject protons 
while keeping as large a fraction as possible of high energy electrons.
It is essential to minimize the trigger rate due to protons, in order not to 
significantly affect the normal PAMELA operation. Moreover, 
the problem of the quantity of mass memory occupied by the events 
acquired in self-trigger mode has to be taken into account. 
Hence, a good event selection already 
at the trigger level is very important.

For these reasons we imposed a threshold of 150~mip on a defined set of planes
and we required that the trigger output from these planes 
occurred within a given time window. Of course, the duration 
of this time window was determined on one hand by the desired sensitivity at 
low energies (i.e., 300 GeV) and on the other hand by the spread in the 
threshold curves. In fact, it is clear that the duration 
of this coincidence should be large enough to include all trigger outputs 
(due to electromagnetic showers) from all the chips involved.

\begin{figure}[ht]
\begin{center}
\includegraphics[height=100mm]{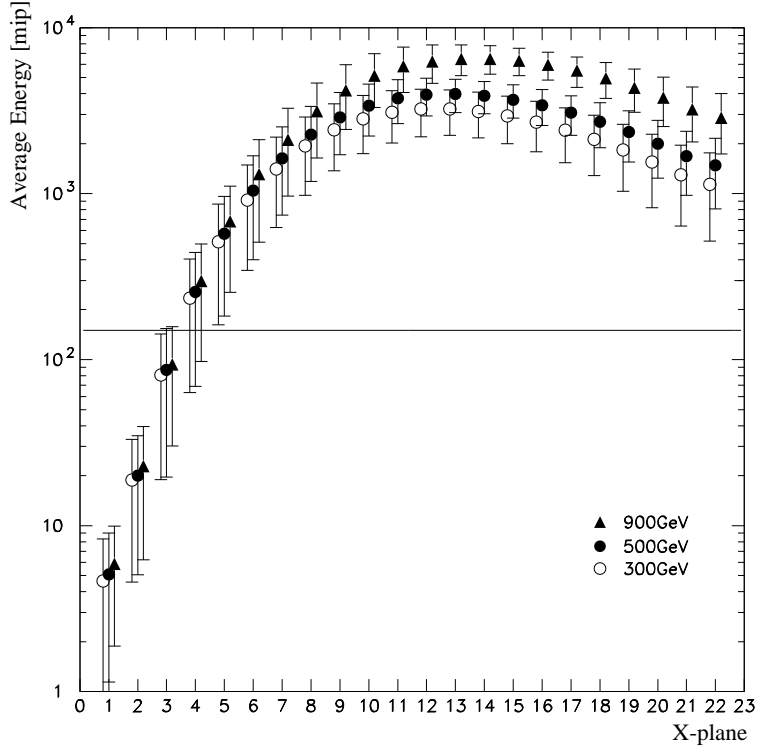}
\end{center}
\caption[]{Average energy detected in the preamplifier reading the highest
energy loss per plane (X views). The solid line indicates the 150~mip 
threshold.}
\label{long}
\end{figure}

As said above, 
the behaviour of the calorimeter in self-trigger mode was studied 
using simulations. 
First of all, we studied the longitudinal development of the 
electromagnetic shower to determine the detected energy in each preamplifier 
of every plane of the calorimeter. 
Figure~\ref{long} shows the average detected energy, with the corresponding 
rms, as a function of the longitudinal depth for the X views 
(the picture for the Y views is similar, just shifted by one plane).
The figure shows only the energy detected in the preamplifier reading the 
highest energy loss in the plane. This 
preamplifier is the first to give an output trigger pulse. From the 
figure it can be seen that, after plane 6, more than 90\% of the electrons 
give an energy loss greater than 150~mip (solid line).
From this result we derived that the best trigger configuration 
is a logic AND between the 6 planes: 
7X, 9X, 11X, 13X, 15X, 17X; 7Y, 9Y, 11Y, 13Y, 15Y, 17Y; 8X, 10X, 12X, 14X, 
16X, 18X; 8Y, 10Y, 12Y, 14Y, 16Y, 18Y. The required AND is actually a 
coincidence within a time window of 100 ns. 

With this configuration, we simulated electrons from 300~GeV to 1~TeV and 
protons impinging also from all sides for energies from 1~GeV to 1~TeV.
The self-trigger response was simulated 
identifying the preamplifier with the highest detected energy in each plane, 
requiring that this energy was greater than 150~mip, converting this energy 
to a time T$_{D}$ using the curve of Figure~\ref{delay2} 
and finally requiring that these 
six T$_{D}$ coincided within 100 ns.
We found that $(99.6 \pm 0.2)$\% of the simulated electrons satisfied the
trigger conditions. To estimate
the proton contamination we took into 
account an experimental proton energy spectrum \cite{boe99}, the 
acceptance of the calorimeter (protons impinging from all sides)
and the fraction of simulated protons 
which passed the trigger conditions. From this calculation we estimated
a trigger frequency due to protons of about 10 mHz, which is 2 orders of
magnitude lower than the expected normal PAMELA acquisition rate.

Considering that the calorimeter in self-trigger condition will work as a
stand alone detector, in combination with a neutron detector, it
will have to identify electrons in a vast background of other particles
(mostly protons) and reconstruct their energy. This was studied with
simulations and is detailed in the next sections.

\subsection{Direction reconstruction}
For a precise determination of the electron energy a good reconstruction of
the electron trajectory is needed. Furthermore, this allows to determine the
amount of material, present around the calorimeter, traversed by the electron 
before entering the detector.

\begin{figure}[ht]
\begin{center}
\includegraphics[height=120mm]{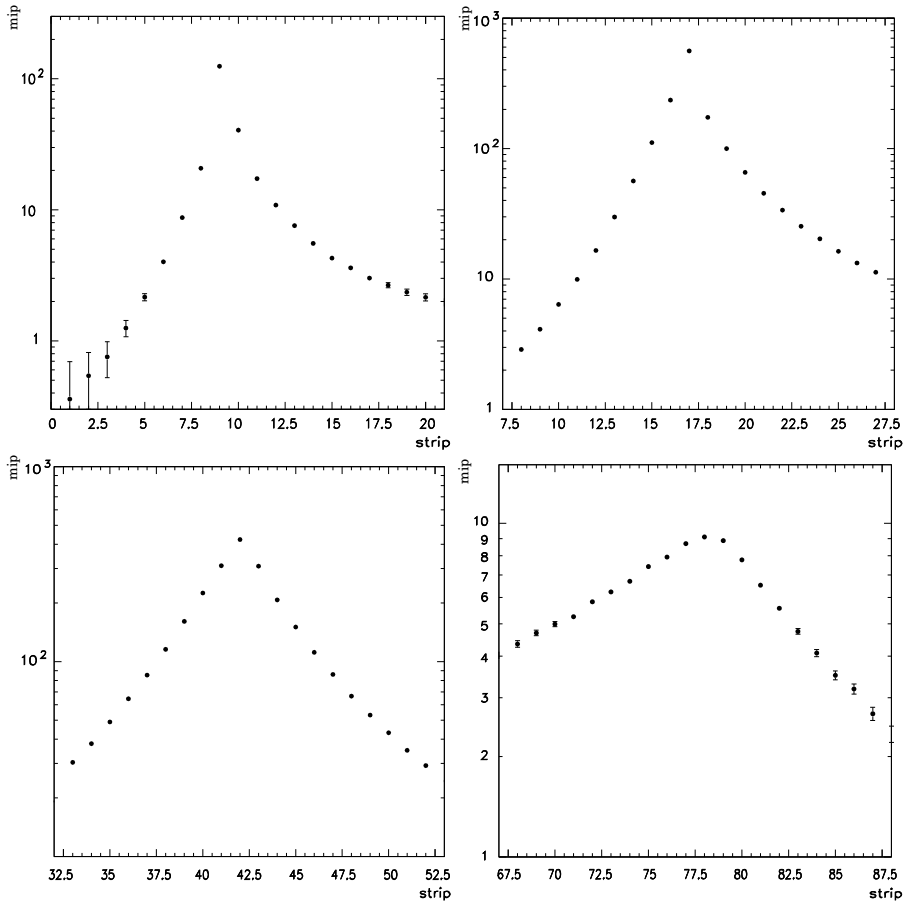}
\end{center}
\caption[]{Average lateral profile of electromagnetic showers induced by 
300~GeV electrons impinging with a zenith inclination of 35$^{\circ}$. From
left to right and from top to bottom the lateral profile
at the third, 
fifth, 10th and 22nd calorimeter plane (X views).}
\label{trans}
\end{figure}

A method based on the centre of 
gravity of the energy losses in each plane for both view was used
to reconstruct the direction.
However, it was found that the simple fit of the centres of gravity
gave biased results
for those electrons having large values of the zenith angle. 
The reconstructed zenith 
angles resulted systematically shifted with respect to the input ones 
by an amount increasing with
the input zenith angle. This can be seen
in Figure~\ref{trans} that shows the lateral profile averaged over
4000 electrons of 300~GeV and input incidence angle of 35$^{\circ}$.
The impact point on the first plane was the same for all the 4000 electrons.
The figure shows the profile at the third, 
fifth, 10th and 22nd calorimeter plane. 
The shower maximum was located close to the 10th plane. It can be observed 
that, except for the 10th plane, the distributions are not symmetrical and, 
consequently, the centre of gravity does not provide the correct position
of the maximum that is located along the electron trajectory.
This asymmetry is due to two effects:
\begin{description}
	\item{1.} for an inclined shower 
	the secondary particles traverse asymmetric amounts of materials;
	\item{2.} the transverse sampling is not orthogonal to the axis
	of inclined shower.
\end{description}
The second, geometrical, effect is more important. These 
effects were studied and a relation was found between the displacements of
the centres of gravity from the shower axis and the incidence angle.
A smaller dependence on the energy of the incident electron was also found 
and it was introduced as a second order correction.
Consequently an iterative procedure was developed to correct for this bias:
\begin{description}
	\item{1.} A first estimation of the angles 
	$\theta_{zx}$ and $\theta_{zy}$
	was obtained fitting with a straight line the centres of gravity
	of the detected energy in each plane around one Moli\`{e}re radius 
	(4 strips) of the cluster of three strips with the highest energy 
	loss. For the fit, the centres of gravity were weighted with the 
	traversal energy loss ($E_{tr}$), i.e.:
\[
	\sigma (x_{i}) \: = \: \frac{x_{i}}{E_{tr}^{0.79}} ,
\]
	with $\sigma (x_{i})$ the error used for the centre of gravity
	($x$) of the $i$ plane ($ i = 1, \ldots, 22$).
	\item{2.} A new set of centres of gravity was calculated using the
	detected energy inside one Moli\`{e}re radius along the fitted
	trajectory.
	\item{3.} Then, these centres of gravity were modified using the 
	functional dependence on the direction and the initial energy
	previously determined. For this correction the trajectory and 
	initial energy (see next section) evaluated at point 1
	were used.
	\item{4.} A new straight line fit was performed using these modified
	centres of gravity.
	\item{5.} The procedure was reiterated back from point 2
	until the reconstructed angles varied less than 1~mrad from
	two consequent iterations.
\end{description}

\begin{figure}[ht]
\begin{center}
\includegraphics[height=100mm]{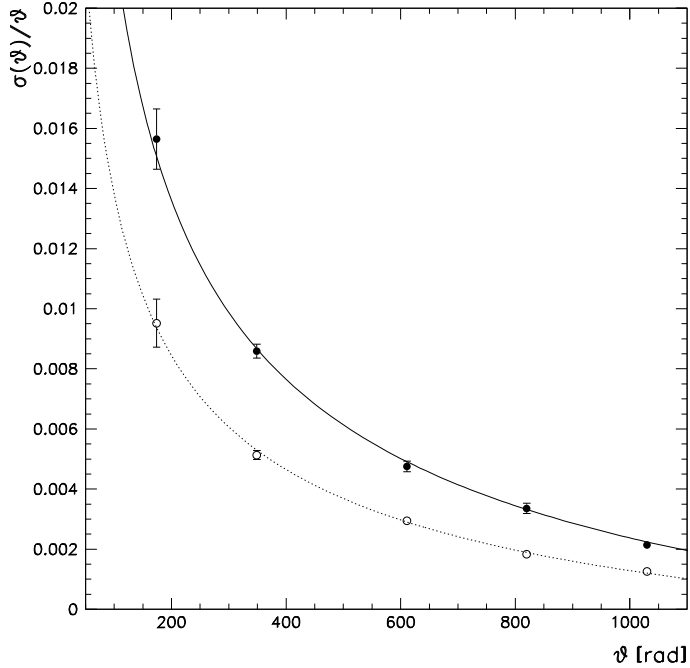}
\end{center}
\caption[]{Relative angular resolution as a function of
input zenith angle for two electron energies: 300~GeV ($\bullet$) and 
1000~GeV 
($\circ$). The two lines are best fits of the data according to 
equation~\ref{eq:resangle}.}
\label{resangle}
\end{figure}

This method permitted to reconstruct the trajectory angle 
from 0 to about 60$^{\circ}$ degrees with a resolution better than 3~mrad. 
Figure~\ref{resangle} shows the 
relative angular resolution as a function of
input zenith angle for two electron energies: 300~GeV ($\bullet$) and 1000~GeV 
($\Box$). The two lines are best fits of the data according to 
equation:
\begin{equation}
\frac{\sigma(\theta)}{\theta} \: = \: \frac{a}{\sqrt{\theta}} - b ,
\label{eq:resangle}
\end{equation} 
where $ a = 0.29 \pm 0.01 $ and $ b = (0.67 \pm 0.06) \cdot 10^{-2}$ at 
300~GeV and
with $ a = 0.183 \pm 0.009 $ and $ b = (0.45 \pm 0.04) \cdot 10^{-2}$ at 
1000~GeV.

\begin{figure}[ht]
\begin{center}
\includegraphics[height=100mm]{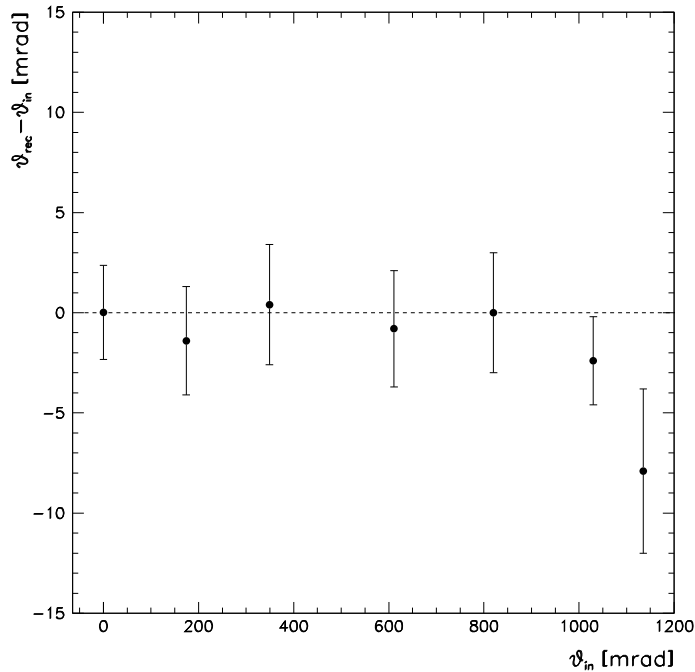}
\end{center}
\caption[]{Difference between the reconstructed angle and the input
zenith angle as a function of zenith angle
for 300~GeV electrons.}
\label{angle}
\end{figure}

The difference between the reconstructed angle and the input
zenith angle is shown in Figure~\ref{angle} as a function of zenith angle
for 300~GeV electrons.
	
\subsection{Energy reconstruction}
\begin{figure}[ht]
\begin{center}
\includegraphics[height=100mm]{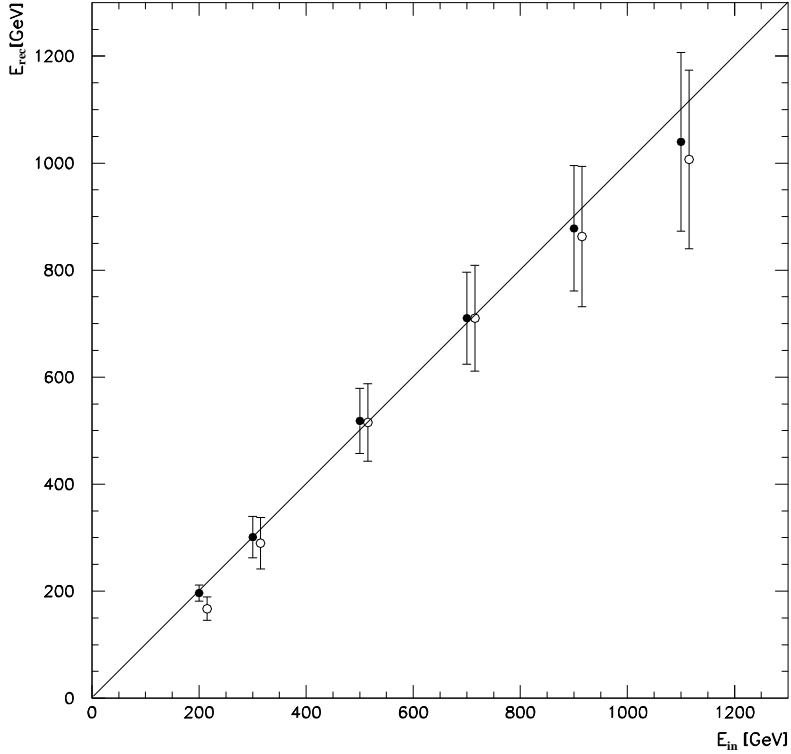}
\end{center}
\caption[]{Reconstructed energy as a function of input energy
for the calorimeter in self-trigger configuration. The solid line
is the identical function. The open circles indicate the total energies 
detected up to three planes after the maximum. The solid circles are the 
total energies corrected for the amount of matter traversed. }
\label{linenergy}
\end{figure}

\begin{figure}[ht]
\begin{center}
\includegraphics[height=100mm]{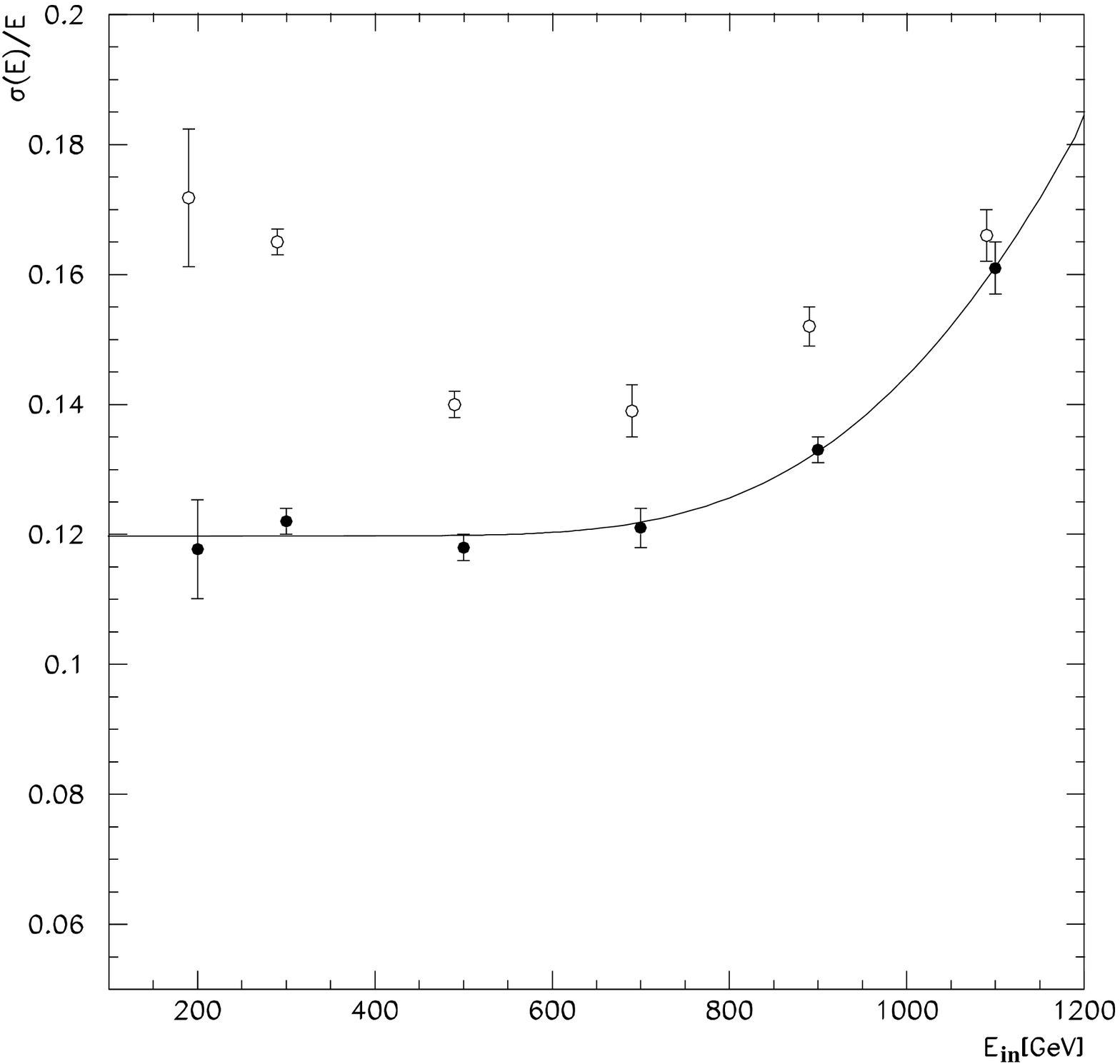}
\end{center}
\caption[]{Energy resolution as a function of input energy for the
energy reconstructed using: ($\circ$)  
the total energies 
detected up to three planes after the maximum and ($\bullet$) the  
total energies corrected for the amount of matter traversed. 
Self-trigger configuration.
\label{resenergy}}
\end{figure}

A large fraction of the energy of electromagnetic showers induced by
high energy electrons (above 100~GeV) leaks out the PAMELA 
calorimeter. This limits the energy reconstruction 
and resolution. Hence, for the energy reconstruction 
in self-trigger mode we used 
the following procedure.
We identified the
plane with the maximum detected energy and summed all the detected energies
up to three planes after this one. In fact, 
up to energies of the order of one TeV the maximum 
of the shower is well contained into the calorimeter also in the self-trigger 
geometry. Then we related this energy with the 
input energy in GeV as it is shown in Figure~\ref{linenergy} (open circles). 
The figure shows that a linear relation exist up to
at least about 1~TeV. The energy resolution obtained with this method is
shown in Figure~\ref{resenergy} (solid circles).

However, the calorimeter, along with a 
longitudinal measurement of the energy loss, 
provides information on the lateral energy. This was
used to improve the energy resolution. In fact, we determined a relation
between the  
energy detected up to the third plane after the one of maximum energy loss 
and the amount of matter traversed, obtained
from the trajectory reconstruction, and we  
used it to correct the 
detected energy. This relation was found to be 
dependent on the incident energy and hence, an iterative 
procedure was used. This procedure was embedded
into the angular reconstruction described in the previous section, hence the
method was reconstructing both the direction and the energy of the incoming
electron. The results from this method are presented 
in Figure~\ref{linenergy} (solid circles), the error bars are one sigma errors
derived from the energy distribution. It can be noticed that this
method provides a 
more precise measurement of the primary energy. 

Furthermore, the 
energy resolution improves as can be seen in Figure~\ref{resenergy}. 
The energy resolution is constant at $\simeq 12\%$ up to about 800~GeV. At
higher energies the resolution decreases because of increasing 
longitudinal leakage and decrease of information on the energy loss
due to saturation of the signal from the strips ($\sim 1100$~mip).  
The resolution was fitted by the following function (solid line in
Figure~\ref{resenergy}):
\begin{equation}
	\frac{\sigma(E)}{E} \: = \: p_{1} \, + p_{2} \cdot 
	e^{-\frac{p_{3}}{E}} ,
\label{eq:resenergy}
\end{equation}
with E kinetic energy in GeV and $p_{1} = (11.9 \pm 0.1) \times 10^{-2}$, 
$p_{2} = 7.6 \pm 3.1$ and $p_{3} = (5.7 \pm 0.4) \times 10^{3}$.

Compared to the energy resolution of the calorimeter in the PAMELA 
normal trigger mode, 
the resolution for the self-trigger configuration is
worse, about 12\% instead of about 5\% at 200~GeV (see section~\ref{s:sim}).
This worsening is due to the different acceptance condition. In the PAMELA 
normal configuration, events impinge nearly orthogonal to the 
first plane of the calorimeter
while in the self-trigger configuration particles
are accepted with zenith angles as large as 60$^{\circ}$ degrees and are
allowed to enter also from the side as long as they cross at least the
fourth plane.

\begin{figure}[ht]
\begin{center}
\includegraphics[height=100mm]{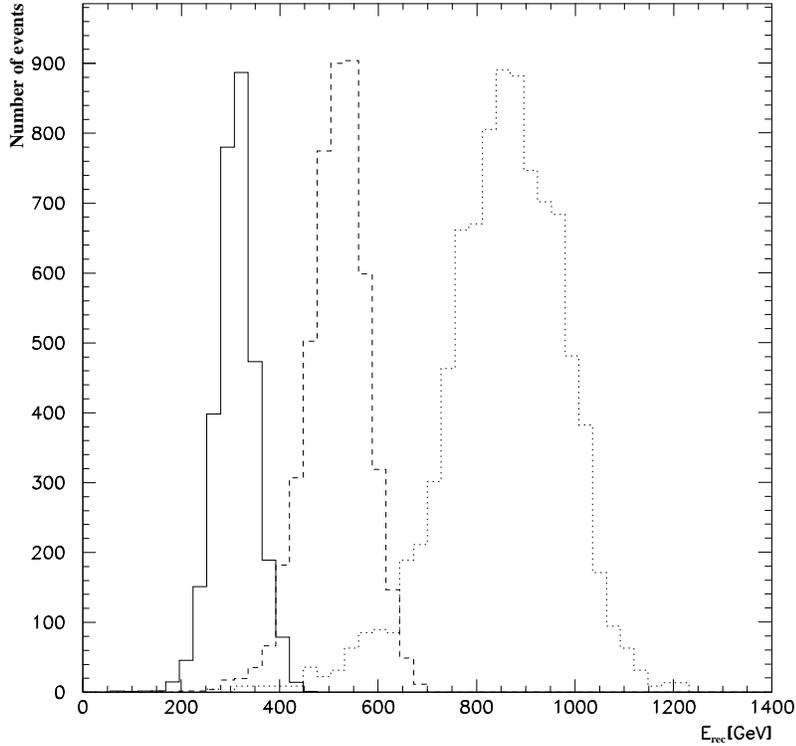}
\end{center}
\caption[]{Distribution 
of the reconstructed energy for three monochromatic electron samples at
300~GeV (solid histogram), 700~GeV (dashed histogram) and 900~GeV (dotted 
histogram). }
\label{endist}
\end{figure}

Figure~\ref{endist} shows the distributions 
of the reconstructed energy for three mo\-no\-chro\-ma\-tic electron samples at
300~GeV (solid histogram), 700~GeV (dashed histogram) and 900~GeV (dotted 
histogram). The distributions can be fitted by 
Gaussian functions which sigmas provide the energy resolution.

\subsection{Particle identification}
In the self-trigger configuration the calorimeter will have to 
reconstruct the energy of the high energy electrons but also identify them
in a vast background of protons and heavier particles. Particle identification
was also studied using simulations. A large number of electrons with energies
between 300 and 1000~GeV and protons with energies from 300~GeV to 3.3~TeV
were simulated. The electrons were simulated with the acceptance of
the self-trigger configuration, while protons were allowed to impinge the
calorimeter also from the lateral sides.
The electron selection criteria were based on the features of the 
electromagnetic showers, as described in section~\ref{s:sim}, 
however without energy-momentum match,
since the calorimeter will work in stand-alone configuration.

The selection efficiency for electrons was found to be 
$(74.7 \pm 0.7)\%$ independent on the energy of the incident electron. The
proton contamination was found to be $(0.15 \pm 0.02)\%$ resulting in an 
proton rejection factor of about 500. Considering the relative abundances 
of protons and electrons in the cosmic radiation, the estimated contamination
flux of protons is of the same order of the electron flux. However, it is
worth reminding that the calorimeter in self-trigger configuration will
work with a neutron counter device that is expected to provide a 
proton rejection factor of about 1000.

\section{Conclusion}
The imaging calorimeter for PAMELA has been designed and 
is presently 
under construction. Laboratory tests and simulations show 
that the instrument can fulfil 
all design requirements for PAMELA.

\ack{We would like to thank Aerostudi Trieste (http://www.aerostudi.it/)
for helping us designing
and constructing the mechanics of the calorimeter. We wish to thank 
Dennis Matveev and Alexandre Pilyar for the help on designing the
electronics and Luigi Vecchiet of Mipot Cormons for technical support
on the assembly of the detectors.}

\end{document}